\documentclass[]{aa}
 \usepackage{epsfig}
\begin{document}
 
 \title{The high-velocity outflow in the proto-planetary nebula Hen 3$-$1475
 \thanks{Based on observations made during service time
 with the 2.5 m Isaac Newton Telescope
 operated on La Palma by the Isaac Newton Group of Telescopes in the
 Spanish Observatorio del Roque de los Muchachos of the Instituto de
 Astrof\'\i sica de Canarias, and observations made with the Hubble 
 Space Telescope, obtained from the
 Data Archive at the Space Telescope
 Science Institute, which is operated by the
 Association of Universities for Research in Astronomy, Inc., under NASA
  contract NAS5-26555 }}

 \author{A.  Riera \inst{1,2}
     \and P. Garc\'\i a-Lario \inst{3}
    \and A. Manchado  \inst{4,5}
    \and M. Bobrowsky \inst{6}
    \and R. Estalella \inst{2}      }
  \institute{Departament de F\'\i sica i Enginyeria Nuclear, Universitat
 Polit\'ecnica de Catalunya, Av. V\' {\i}ctor Balaguer s/n, E-08800
 Vilanova i la Geltr\'u, Spain 
             \email{angels.riera@upc.es}
         \and
      Departament d'Astronomia i Meteorologia, Universitat de
 Barcelona, Av.\ Diagonal 647, E-08028 Barcelona, Spain 
        \and 
 ISO Data Centre. Science Operations and Data Systems Division. Research and
 Scientific Support Department of ESA. Villafranca del Castillo. 
 Apartado de Correos 50727. E-28080 Madrid, Spain 
      \and 
 Instituto de Astrof\'\i sica de Canarias, E-38200 La Laguna
 (Tenerife), Spain   
 \and     
Consejo Superior de Investigaciones Cient\'\i ficas, Spain 
 \and
 Challenger Center for Space Science Education. 1250 North Pitt Street,
 Alexandria, VA 22314, USA 
 }
 \offprints{A. Riera}
 \date{Received October 30, 2002; accepted February 13, 2003}
 \authorrunning{Riera et al. }
 \titlerunning{High-velocity outflow in Hen~3$-$1475}
 
 \abstract{
 The proto-planetary nebula Hen~3$-$1475 shows a remarkable highly
 collimated optical jet with an S-shaped string of three pairs of knots
 and extremely high velocities. We present here a detailed analysis of the 
 overall morphology, kinematic structure and the excitation conditions of 
 these knots based on deep ground-based high dispersion spectroscopy 
 complemented  with high spatial resolution spectroscopy obtained with 
 STIS onboard HST, and WFPC2 [N~II] images.  
 The spectra obtained show double-peaked, extremely wide emission line 
 profiles, and a decrease of the radial velocities with distance to the source 
 in a step-like fashion. We find that the emission line ratios observed in the
 intermediate knots  are consistent with a spectrum arising from
the recombination region of a shock wave with shock velocities ranging from 
100 to 150 km~s$^{-1}$.  
 We propose that the ejection velocity is varying as a function of time 
with a quasi-periodic variability (with timescale of the order of 100 years)
 and the direction of ejection is also varying with a precession period 
of the order of 1500 years. Some slowing down with distance along the axis 
of the Hen~3$-$1475 jet may be due to the entrainment process and/or 
to the enviromental drag. This scenario is supported by geometric and 
kinematic evidence: firstly, the decrease of the radial velocities along
 the Hen~3$-$1475  jet in a step like fashion; secondly, the kinematic 
structure observed in the knots; thirdly, the point-symmetric morphology 
together with the high proper motions shown by several knots;  
and finally the fact that the shock velocity predicted from the 
observed spectra of the shocked knots is much slower than the velocities 
at which these knots move outwards with respect to the central source.       
}

\maketitle

 \section{Introduction}
 
 Hen 3$-$1475 is a highly collimated bipolar proto-planetary nebula (PPN,
 hereafter), which displays a spectacular S-shaped string of point-symmetric
 knots extending over $17''$ along its main axis. 
 It was first observed from the ground by  Riera et al. (1995, hereafter R95) 
 and from the ground and space by Bobrowsky et al. (1995)
 and by Ueta et al. (2000),  and identified as a massive post-AGB star
 rapidly evolving into the planetary nebula (PN, hereafter) phase.
 More recently, with the help of HST WFPC2 observations, it has been 
 possible to resolve the subarcsecond structure of the outer
 knots and of the inner region of this extremely complex source 
 (Borkowski et al. 1997). 
 
 Hen 3$-$1475 was recognized for the first time as a transition object in
 the post-AGB phase by Parthasarathy \& Pottasch (1989) from the analysis of
 its far infrared IRAS colours. Later, it was revealed as a
 highly collimated bipolar PPN by R95, showing the  highest outflow
 velocity ever observed for a PPN and a peculiar 
 point-symmetric morphology consisting of
 pairs of knots symmetrically distributed with respect to the central star.
 From the observed spectra, together with the
 detection of highly supersonic velocities, it was concluded that the emission
 in the outer knots was produced in a shock which propagates
 through a nitrogen-enriched medium (R95).
 
 A velocity gradient decreasing from
 900 km~s$^{-1}$ in the innermost region close to the central star down to
 450 km~s$^{-1}$ in the outer knots was also found. This 
 was interpreted as a sign of variability of the ejection velocity by R95
 and Riera et al. (2001,2002). However, Borkowski \& Harrington (2001; BH01,
 hereafter) interpret
 this velocity gradient in a different way, as a sign of efficient
 deceleration of the jet by a much slower bipolar outflow. BH01 also suggest that
 the wide line profiles observed in the knots are a sign of violent deceleration 
 of the jet. 

 An inclination angle of the bipolar outflow of $50^{\circ}$ with respect to the plane 
of the sky, was derived by BH01 from the Doppler shifts 
of the scattered stellar H$\alpha$ line, considering that the reflected light is redshifted 
due to the motion of the dust away from the star, and that the H$\alpha$ emission is also   
red- or blueshifted as a consequence of the inclination of the flow.

 The strong far infrared excess detected by IRAS can be
 identified as the remmant emission from the dust grains formed in the
 circumstellar shell during the previous strong mass loss AGB phase,
 while the current presence of strong P-Cygni profiles in the Balmer lines
 indicates that mass loss is still on-going (although at a much lower rate).
 This is confirmed by the detection of thermal emission from  hot dust  in
 the near infrared  (Garc\'\i a-Lario et al. 1997). 
 
 The nitrogen overabundance derived from the optical spectroscopy
 (R95) and the high luminosity ($> 10^4~ L_{\odot}$ at a distance
 of ~5.8 kpc - as explained below) suggest 
 a high mass progenitor for the central 
 star of Hen 3$-$1475.
 
S\'anchez Contreras \& Sahai (2001) have recently analysed two blue-shifted 
 absortion features detected in the light of H$\alpha$ arising
from a region very close to the central source using STIS data. 
 They detect the presence of 
 two different winds both with a kinematic age of only tens of years: a fast
wind (flowing with velocities $\sim$ 150 --- 1200 km~s$^{-1}$) 
 and an `ultrafast' wind  which is flowing with velocities 
 up to 2300 km~s$^{-1}$. S\'anchez Contreras \& Sahai (2001) find that the 
 `ultrafast' wind is highly collimated (at distances $\sim$ 10$^{16}$ cm) 
 in a direction which differs from previous mass ejection axes 
 and shows a large velocity gradient. This `ultrafast' wind is identified as
 a `pristine' post-AGB outflow which has not yet been altered by its 
 interaction with the AGB envelope.
 
VLA observations revealed  the presence of a compact radio
 continuum source close to the central star (Bobrowsky et al. 1995,
 Knapp et al. 1995), indicating that the central star has already become
 hot enough to ionize the gas in its vicinity.

The presence of a large amount of neutral gas in the envelope has also 
 been detected with the help of observations in the light of the CO molecule emission 
 as an expanding torus of material (Knapp et al. 1995, 
 Bujarrabal et al. 2001). In addition, Hen 3$-$1475  exhibits
 strong OH maser emission (te Lintel Hekkert 1991; Bobrowsky et al. 1995).
Both CO and OH emission show a broad (60 km~s$^{-1}$) profile
 indicative of a fast molecular outflow which appears to be extended
  ($2''$). Bujarrabal et al. (2001) identify in the wide CO profile two
 different components: a slow component (with an expansion
  of $\sim7$ km~s$^{-1}$) and a fast outflow, which 
 is assumed to be bipolar in the direction of the nebular axis. 
 Similarly, the OH emission profile consists of a plateau of emission
 and sharp peaks at $+21.5$ and $+70.9$ km~s$^{-1}$, implying an
 expansion velocity of at least 25 km~s$^{-1}$ in the outer parts of the
 neutral envelope. The blueshifted and
 redshifted peaks are displaced $1''$ in the direction of the optical jet,
 which suggest an aspherical shell. It is remarkable that the OH radial
 velocity increases linearly with distance to the source, a situation that has
hen.graphic been observed in other PPNe, and can be explained in terms of bipolar 
 winds (Zijlstra et al. 2001).  

  In this paper we present a detailed analysis of the kinematics of the
 high-velocity outflow  based on new ground-based
 spectroscopic data taken in the wavelength range corresponding to the
[O {\sc i}] 6300 \AA, [N {\sc ii}] 6548, 6584  \AA, H$\alpha$, and 
 [S {\sc ii}] 6717, 6731 \AA~
 emission lines, covering $\pm9''$ \rm{NW} and SE of the central star.
 This is complemented with the high spatial resolution 
 information provided by the Space Telescope Imaging Spectrograph (STIS) data.
 Based on this new
 comprehensive data set we re-discuss the overall kinematics and the excitation
 of the knots observed in Hen~3$-$1475.
 
In  Sect. 2 we describe the observations on which our analysis is based,
 while in Sect.  3 we present the improved results
 obtained on the morphology description and the kinematic properties of the  
 knots. Possible excitation mechanisms for the lines observed and 
 several different scenarios which could be responsible for the complex high velocity 
 field associated to Hen 3$-$1475 are  discussed in Sect. 4 in connection 
 with the nature and evolutionary stage of the central source, 
 while our main conclusions are presented in Sect.  5.

 \section{Observations}
 
 \subsection{HST images}
 
 Two sets of high spatial resolution optical WFPC2 images of 
 Hen~3$-$1475 taken at different epochs with HST on 1996 June 6 and
 1999 September 19 were used to study the overall morphology of the source and 
 to measure the proper motion of the knots.  The images
  were obtained in both cases with the Planetary Camera  (800$\times$800 pixels
  with a plate scale of $0\farcs0455$ pixel$^{-1}$) and through
  the F658N narrow filter ($\lambda_c$ = 6590 \AA;  $\Delta \lambda$= 28.5 \AA).
  The images taken in 1996 were originally  part of  Cycle 6
  proposal 6347 (P.I.:  K.F. Borkowski), while the 1999 images were part of
   Cycle 7 proposal 7285 (P.I.: J.P. Harrington).
  These images went through the pipeline processing at the Space 
  Telescope Science Institute. After  HST pipeline calibration the 
  cosmic rays were removed using the ``crrej'' task within the IRAF\footnote{IRAF is 
 distributed by the National Optical Observatories, which 
  is operated by the Association of Universities for Research in Astronomy, 
  Inc., under contract with the National Science Foundation.}
   package.

  \subsection{High resolution optical spectroscopy}

  Two complementary spectroscopic data sets are combined in 
  this work: (i) deep ground-based long slit optical 
  spectroscopy along the bipolar axis with high 
  spectral resolution (R$\sim$20,000) 
  and (ii) spectroscopy with very
  high spatial resolution along the bipolar axis obtained with STIS 
  (Space Telescope Imaging Spectrograph) onboard HST.

  \subsubsection{Ground-based spectroscopy}
  
  The ground-based long slit optical spectra of Hen~3$-$1475 were 
  obtained in June 1998 as part of the Isaac Newton Group service program
  using the Intermediate Dispersion Spectrograph (IDS) equipped with
the 500 mm camera and a TEK CCD installed at the Cassegrain focus of the 2.5 m
  (f/15) Isaac Newton Telescope  at the Observatorio del Roque de los Muchachos
  (La Palma, Spain). The slit width used was $2''$,  oriented along
  the major axis of the bipolar nebula passing through the central star at a
  P.A. of  $135^{\circ}$, covering not only the innermost and intermediate 
  knots but also grazing the outer knots. 
  
   Two exposures, each one covering a different spectral range,
  were performed with the high resolution grating H1800V
  centered at 6640 \AA~ and at 6300 \AA~ respectively,
  yielding a spectral sampling of 0.46 \AA,  equivalent to a velocity 
  resolution of 21 km~s$^{-1}$. The spatial sampling is 
  $0\farcs33$ pixel$^{-1}$ in both cases. With the first exposure
  ($\lambda_c$ = 6640 \AA) we covered the [N~{\sc ii}] 6548, 6584 \AA,
  H$\alpha$ and [S~{\sc ii}] 6717, 6731 \AA~ emission lines,  while the second 
  exposure ($\lambda_c$ = 6300 \AA) was intended to cover  the [O~{\sc i}] 6300
  \AA~ emission line, which is of great importance in determining the
  excitation mechanisms involved. The exposure times were 900 and 1800 s
  for the first and second exposures, respectively, improving the S/N ratio of 
  previous observations (R95)  
  and extending the observations up to a distance of $\pm$ $9\farcs0$ from the 
  central star. This meant that, for the first time,  we were able to study the
  kinematic structure of the complete outflow.
  
    Individual spectra were corrected for bias, flat-field and cosmic
 ray events, and were calibrated in wavelength using the standard tasks 
for long-slit spectroscopy within the IRAF package. 
The spectra were not flux calibrated. 
  
  \subsubsection{HST spectroscopy}
  
   STIS data were retrieved from the HST Data 
  Archive. Long-slit spectra were obtained with the STIS slit oriented 
  along the bipolar axis of Hen~3$-$1475 on 1999 June 15-16. 
  They were originally part of Cycle 7 proposal 
  7285 (P.I.: J.P. Harrington). The spectrograph slit had a projected 
  length of $52''$.  The G750M grating, centered at $\lambda_c$ = 6581 \AA,
  was used, covering the wavelength region including [O~{\sc i}] 6300 \AA,
  [N~{\sc ii}] 6548, 6584 \AA, H$\alpha$ and [S~{\sc ii}] 6717, 6731 \AA.
   A $0\farcs1$ slit width was used, giving a spectral resolution of 
  $\sim50$ km~s$^{-1}$ and a dispersion of 0.56 \AA~pixel$^{-1}$. The spatial 
  sampling was $0\farcs05$.

  Spectra were obtained at nine parallel positions with increasing offsets
  with respect to the central star of
  $0\farcs1$ at both sides of the nebula from $-0\farcs4$ to $+0\farcs4$. 
  The central spectra (offsets $-0\farcs1$, $0\farcs0$, and
  $+0\farcs1$) cover the bright innermost regions (including the knots 
  named \rm{NW1} and \rm{SE1}; see Fig.~\ref{FigHST}). The spectra with  
  $+\farcs2$ to  $+\farcs4$ displacements from the central star position 
  partially covered the \rm{SE2} knot with the slit at its western edge, but the
  blue-shifted \rm{NW2} knot was unfortunately missed (see Fig.~\ref{FigHST}).  
 They all were calibrated at the Space Telescope Science Institute following the 
  HST pipeline calibration (see BH01 for a detailed description).
  
   Other authors have previously reported in the literature some of the
  results derived from the analysis of this STIS spectroscopy. BH01 had 
  their main interest 
  in the study of the scattered emission in order to determine the 
  inclination of the jet axis. Based on the same data set,  
  S\'anchez Contreras \& Sahai (2001) have recently  
  presented a detailed study of the high velocity wind from the central star
  at the innermost regions of the nebula. However, 
  the STIS spectra arising from the knots have not yet been discussed in 
  detail. As we will discuss later, long-slit STIS spectra  
  confirm previous results reported in the literature but also contain 
  interesting new information about the high-velocity bipolar outflow, never
  discussed before,  illustrating the changes in the line profiles 
  and radial velocities along the flow with unprecedent spatial resolution.

  \section{Results}

  \subsection{Morphology of the outflow}
  
  The images taken with HST using the narrow F658N filter esentially trace
  the ionized gas in Hen 3$-$1475 through the detection of
  the nebular emission coming from the [N{\sc ii}] 6584 \AA~ line. However, it
  should be noted that  losses are expected in the detection of this
  line from the more extremely blueshifted regions of the nebula, not
  covered by the filter, due to the extremely large velocities involved. 
  Moreover, extremely redshifted H$\alpha$ emission
  is also expected to contribute to the emission detected through this
  filter in the inner regions of the redshifted lobe, making more difficult a 
  straightforward interpretation of the surface brightness distribution 
  observed in the HST images.
  
     \begin{figure*}
     \centering
     \includegraphics{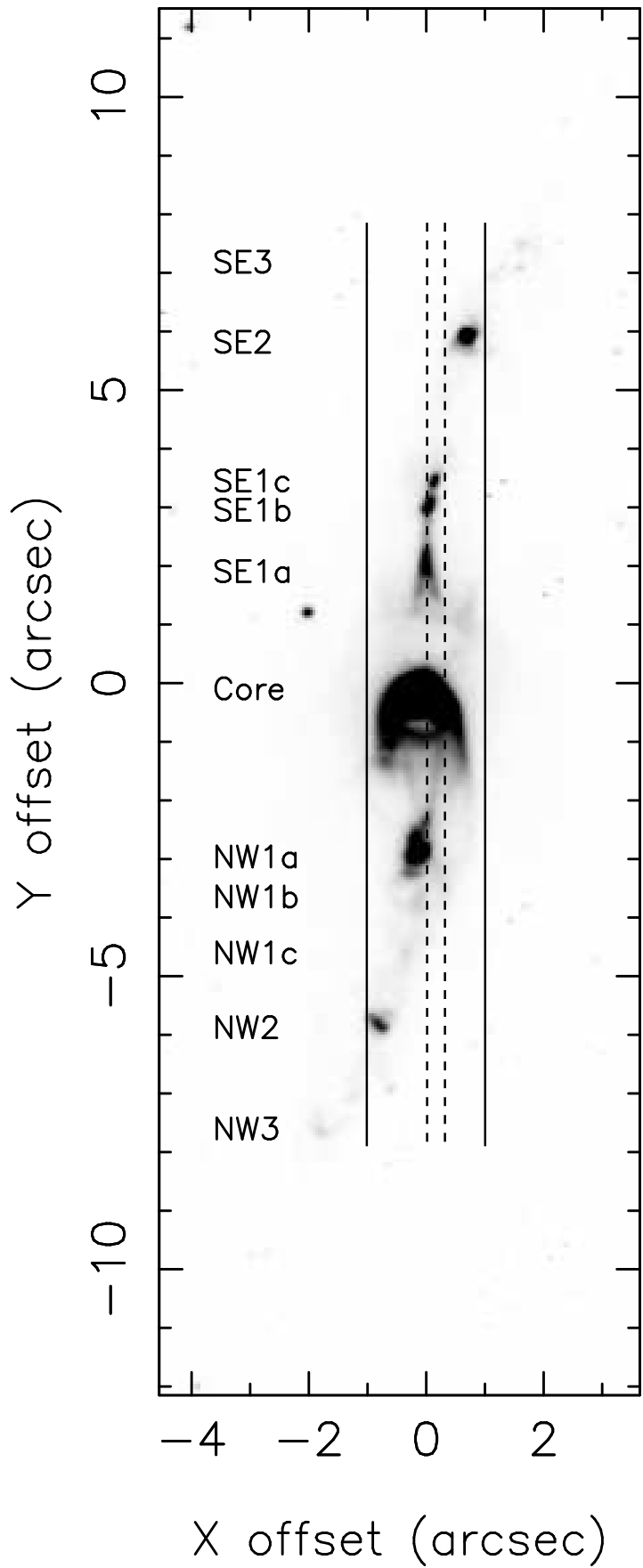}
     \caption{HST image of Hen 3$-$1475 taken through the F658N filter which 
  esentially traces the emission from the [N {\sc ii}] 6584 \AA, derived from 
  the 1999 observations.
  The (0,0) position was set at the location of the core, where the maximum 
  emission was measured. The slit position used for the ground-based 
observations (with  a width of $2''$) is shown by two solid lines.    
Dashed lines show the STIS slit positions for the central spectra (offset 
$0\farcs0$) and for an offset of $0\farcs3$ (see text).}
                \label{FigHST}
      \end{figure*}  
  
   In Fig.~\ref{FigHST} 
  we show as an example one of the F658N images taken in 1999 where
  we can see the jet-like structure of Hen 3$-$1475
  and  three pairs of  symmetric emission knots, named following the
  nomenclature used by Borkowski et al. (1997). 
  The overall shape looks very similar to that observed in the images
  taken in 1996.  From these images Borkowski et al. (1997) 
  remarked on the presence of three pairs of 
 well-collimated  knots
  and the presence of converging funnel-shaped structures
  connecting the innermost knots and the core region in the \rm{NW} and SE lobes, 
  which are still clearly visible in the 1999 images.
  
  The knots \rm{SE1} and \rm{NW1} show subarcsecond structure,
  with the existence of three well defined compact subcondensations which are
  labelled with a letter after the number (see Fig.~\ref{FigHST}). The three 
  subcondensations of the \rm{SE1} knot show similar intensities in the 
  emission line images and are located at distances of $2\farcs27$, 
  $3\farcs29$ and $3\farcs65$ from the central core. In the \rm{NW} knot  the 
  subcondensations \rm{NW1b}  and \rm{NW1c}  are much fainter than \rm{NW1a}  in the 
  emission line images (see Fig.~\ref{FigHST}). However,  these  
  substructures are much  more prominent in  scattered light (Riera et al. 
  2001).  Subcondensations NW1a, \rm{NW1b}  and \rm{NW1c}  are located 
  at larger angular distances from the central core than their SE counterparts. 
   \rm{NW1a}  is located at $2\farcs60$ from the central source, while the \rm{NW1b}  and 
   \rm{NW1c}  are located at $3\farcs46$ and $4\farcs36$, respectively. This is 
  most probably a projection effect. Assuming that the location of the \rm{NW1} and \rm{SE1} 
 knots is symmetric with respect to the central source, the difference between the 
 projected distances for each subcondensation and its counterpart, can be explained  
 if the jet is not coplanar. In this case, a difference 
$\Delta i= i_{NW} - i_{SE} < 10^{\circ}$, where  $i_{NW}$ and  $i_{SE}$ are the 
 inclination angles of the \rm{NW} and \rm{SE} sides of the jet, would explain the discrepancy 
between the projected distances from the central source to the \rm{NW1} subcondensations 
 and the \rm{SE1} counterparts. 
  
  As it was  already proposed in R95, 
 the S-shaped morphology of the jet of Hen 3$-$1475 suggests that the direction
  of ejection is not constant. If these variations are interpreted as a 
  consequence of precession, 
  the observed morphology can be fitted with a precession angle 
   $<$ 10$^{\circ}$ and a period of $\sim1500$ years (as quoted by R95).

  \subsection{Determination of the systemic velocity}
  
  The systemic velocity of Hen 3$-$1475 can be estimated assuming that 
 the velocity field of the knots is symmetric about the central source, 
so that the velocity is the same for the redshifted and blueshifted knots of 
 each pair.  This way, from our IDS ground-based spectra, we 
  derive the value of $+44$ km~s$^{-1}$ both from the outermost
   (\rm{NW3} and \rm{SE3}) and from the middle (\rm{NW2} and \rm{SE2}) knots. This velocity is 
  consistent within the errors with the value of $+40$ km~s$^{-1}$ 
   reported by BH01 from the circumstellar FeII emission lines and it 
  is also close to the velocity centroid of the OH 1667 MHz emission, 
  which is $+46$ km~s$^{-1}$ (Bobrowsky et al. 1995; Zijlstra et al. 2001) 
  and the systemic velocity of $+$48 km~s$^{-1}$ determined from the CO 
  profiles (Bujarrabal et al. 2001). 
  In the following we will adopt a local standard of rest (LSR) systemic velocity 
of $+44$ km~s$^{-1}$.

  \subsection{Kinematics of the outflow}

   The spatially resolved kinematic information contained 
  in our ground-based spectroscopy, combined with the STIS observations,
  has been used to fully 
  map for the first time the
  velocity of the gas along the bipolar axis of Hen 3$-$1475 over the whole
  extension of the nebula.  
 
 For this, we have followed the standard method
  of fitting multiple gaussians to the observed emission line profiles 
  observed in our ground-based spectroscopy 
  (fitting was done using standard IRAF packages) and transforming the 
 derived central wavelengths of each individual line into LSR velocities. 
  
   \begin{figure*}
     \centering
     \includegraphics{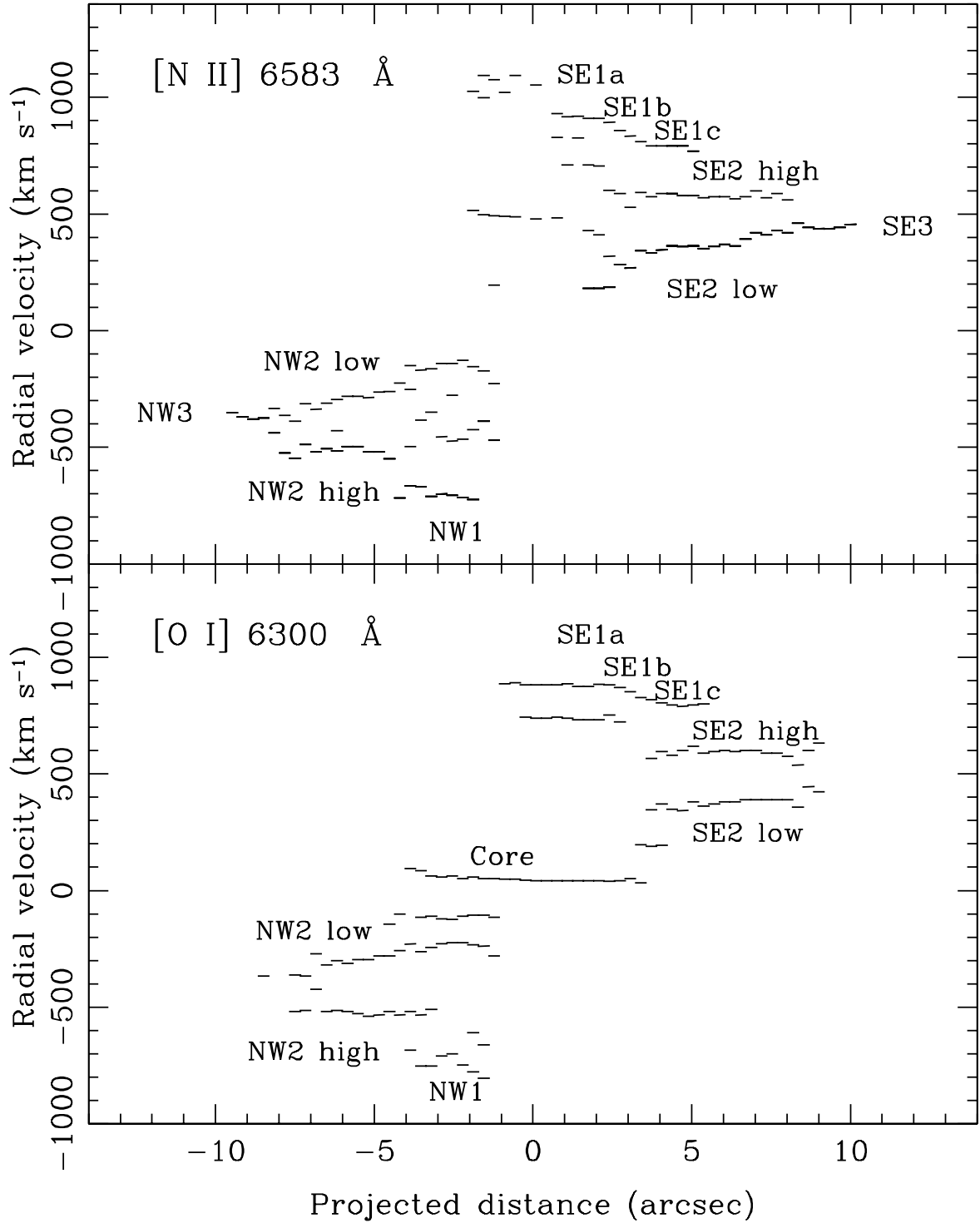}   
     \caption{Relative radial velocity {\it vs.} distance to the central 
  source along the axis of the nebula for  [N~{\sc ii}] 6584 \AA~ ({\it upper panel}) 
  and  [O~{\sc i}] 6300 \AA~ ({\it lower panel}). 
  Relative velocities are quoted
  with respect to the systemic radial velocity (= 44 km/s). 
   The labels identify the position of the different components of the knots. 
  The `low' and `high' levels identify the low and high-velocity components 
  shown by the intermediate knots. In the lower panel the label `core' 
accounts for the  scattered [O~{\sc i}] emission arising from the central core 
  }
                \label{FigVel} 
      \end{figure*}

  These fits have been obtained for the [N~{\sc ii}] 6584, 6584 \AA, 
H$\alpha$, [S~{\sc ii}] 6717,6731 \AA~  and [O~{\sc i}] 6300 \AA~ emission 
  lines. Fig.~\ref{FigVel} shows the position-velocity diagrams derived 
  from [N~{\sc ii}] 6584 \AA, and [O~{\sc i}] 6300 \AA~ 
  emission lines. The results obtained using different 
forbidden emission lines are consistent each other 
  within the errors and they are all also very similar to
  those obtained in the light of H$\alpha$ (not shown in Fig.~\ref{FigVel}). 
  The position-velocity diagrams of H$\alpha$ and [S~{\sc ii}] 6717 \AA~ were 
  shown in Riera et al. (2001, 2002).
  
  Note that the [O~{\sc i}] emission lines observed in the central
  region show very low associated velocities which correspond to reflected 
  emission originated in the central region which is being scattered by 
  dust particles (labeled as `core' in  Fig.~\ref{FigVel}).  
The kinematic pattern of the scattered light, also observed
  in H$\alpha$, has already been discussed by Riera et al. (2001)
  (based on ground-based spectra) and BH01 (based on the STIS spectra), and 
  will not be further discussed  here. 
   In all these diagrams the position of the central source is identified
  with the centroid of the spatial distribution (along the slit) of the
  H$\alpha$ and [O~{\sc i}] 6300 \AA~ emission, which is located at $0\farcs0$.

\begin{figure*}
     \centering
     \includegraphics[width=14cm]{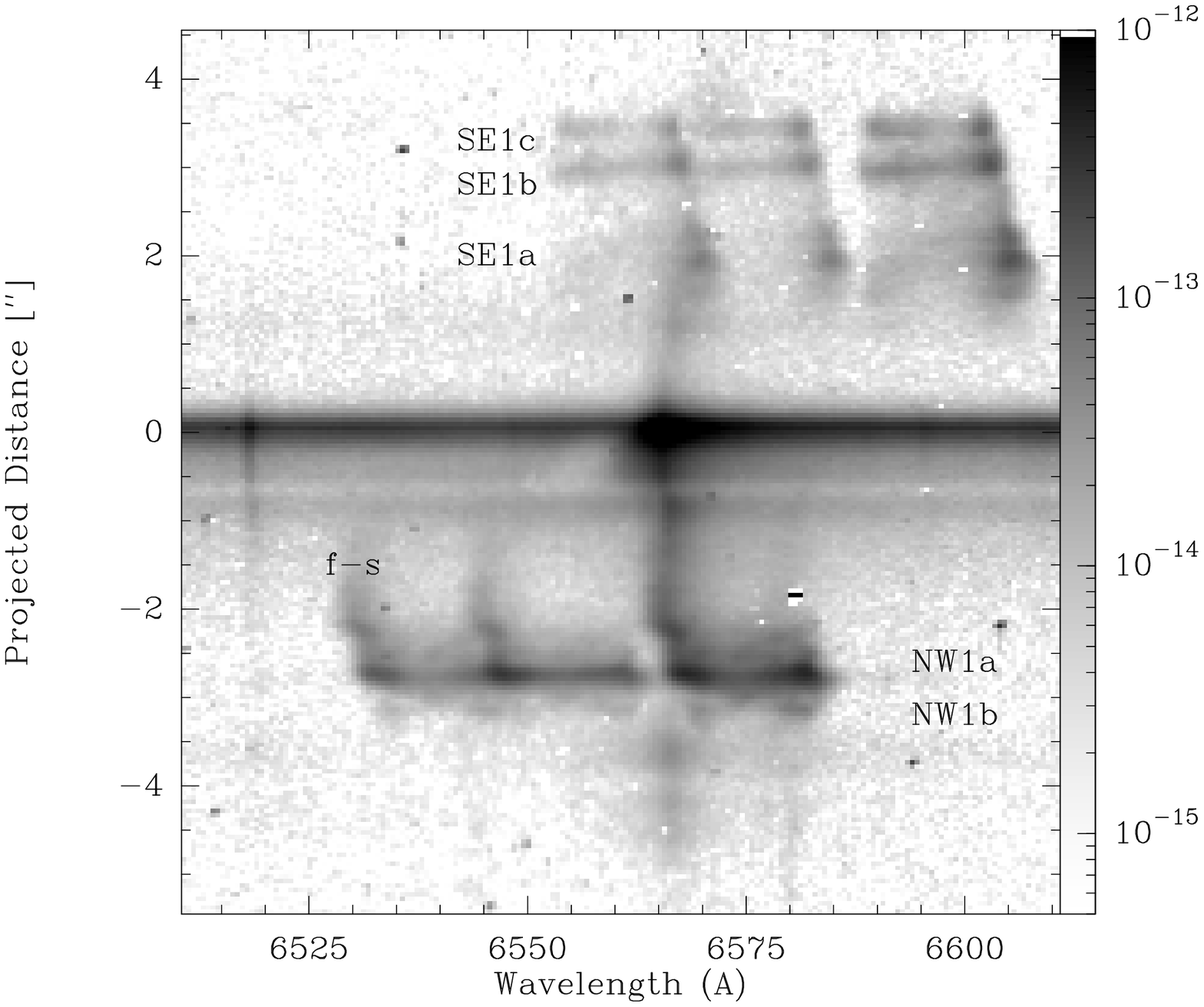}
     \caption{Position-velocity diagram for [N~{\sc ii}] 6548,
  6584 \AA, and H$\alpha$ 6563 \AA~ obtained with  STIS 
  at slit positions with
  offsets $-0\farcs1$, $0\farcs0$ and $+0\farcs1$ with respect to the central
  source.  We have used a logarithmic grey scale. The grey bar indicates the 
 flux value in erg cm$^{-2}$ s$^{-1}$ \AA$^{-1}$. 
  The ordinate gives the position along the spectrograph slit. The main 
features have been identified with a label. The structures labelled as ``f-s'' are 
the funnel-shaped features (see text). 
  }
                \label{FigSTIS2}
      \end{figure*}

In Fig.~\ref{FigSTIS2} we show 
the kinematic structure of the innermost regions of the nebula derived from 
 the STIS measurements corresponding to slit positions along the bipolar
 axis at angular distances of $-0\farcs1$, $0\farcs0$, and $+0\farcs1$ from  
 the central source in the light of [N~{\sc ii}] 6548, 6584 \AA~ and H$\alpha$.
  These observations covered completely the innermost \rm{SE1} and \rm{NW1} knots 
(see Fig.~\ref{FigHST}).
  Extraordinary line widths, up to $\simeq1050$ km~s$^{-1}$, are clearly
  visible in the position-velocity diagram of these knots along with remarkable 
  double-peaked profiles.

   \begin{figure*}
     \centering
     \includegraphics[width=14cm]{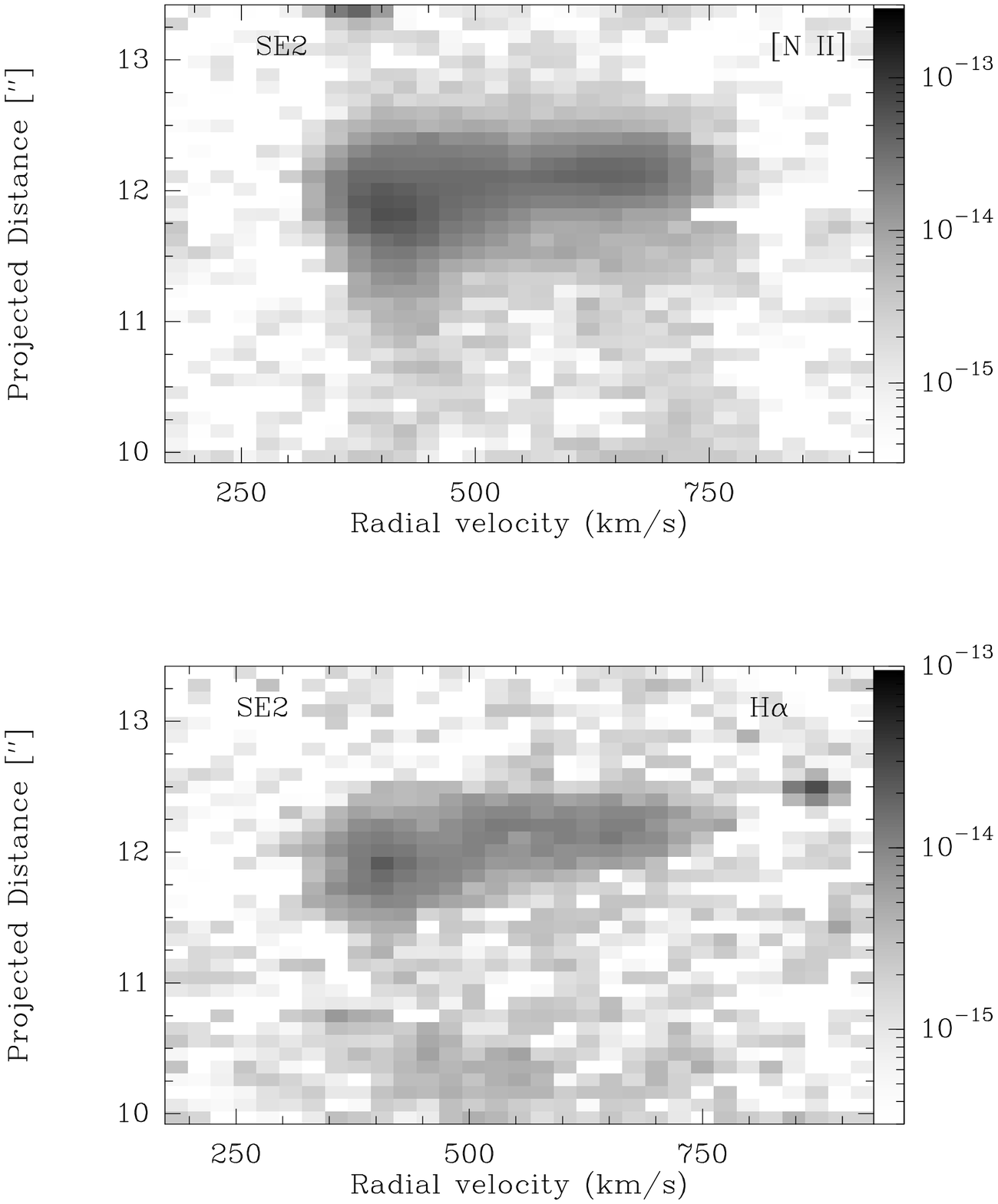}
     \caption{Position-velocity diagram for [N~{\sc ii}] 6584 \AA~ (upper panel) and
  H$\alpha$ 6563 \AA~ (lower panel) obtained with STIS (at slit positions $+0\farcs2$,  
  $+0\farcs3$, and $+0\farcs4$  with respect to the central source)  
through the red-shifted knot \rm{SE2}. We have used a 
grey logarithmic scale. The grey bars indicate the 
 flux value in erg cm$^{-2}$ s$^{-1}$ \AA$^{-1}$.  The ordinate gives the position along the 
spectrograph slit. 
  }
                \label{FigSTIS1}
      \end{figure*}

  Similarly, in Fig.~\ref{FigSTIS1} we show in detail the sub-structure of 
  the \rm{SE2} knot as observed with STIS.
  In this case, the velocity-position diagram 
  shows the combined information extracted from the exposures taken with the
  slit positions  $+0\farcs2$, $+0\farcs3$, and $+0\farcs4$ in the light
  of  H$\alpha$ and [N~{\sc ii}] 6584 \AA~ only. 
  Again we can see clearly the presence of remarkable double-peaked profiles, 
  which were previously reported by R95.
  
   Fig.~\ref{FigSTIS2} shows weak emission in the region 
  between the core and \rm{NW1a}  at a distance of $1\farcs8$ from the 
central source which corresponds to the converging funnel-shaped structures
  connecting the innermost knots and the core region seen in the 
  emission line images (see Fig.~\ref{FigHST}). 
There is some contribution of scattered light from 
  the strong H$\alpha$ emission from the central source, but part of the 
  observed emission is intrinsic to the jet. The emission lines formed in 
  this region show asymmetric broad single-peaked emission line profiles, 
  moving at $-920$ km~s$^{-1}$ with respect to the systemic velocity and 
  with a redward tail up to $-500$ km~s$^{-1}$.
  
  One of the most remarkable results found in the analysis of this outflow is 
  the decrease in radial velocity when moving outwards along the jet. 
  Within the knot NW1, we observe an abrupt 
  decrease of $145$ km~s$^{-1}$ in just $0\farcs875$ (see Fig.~\ref{FigSTIS2}). 
 The high-velocity peak of \rm{NW1a}  (at $2\farcs6$ from the central source) 
 occurs at $-766$ km~s$^{-1}$, while at  $3\farcs2$ from the central 
source (close to  NW1b) the emission peak is at $-720$  km~s$^{-1}$.  

 We also see rapid velocity changes within \rm{SE1} in the transition regions
  from each subcondensation to the next one. The strongest emission occurs 
in the high-velocity component, which is centered at $\simeq +890$  
(SE1a) to $+740$ (SE1c) km~s$^{-1}$.   The radial velocity decreases outwards 
  in a step-like fashion, with an abrupt decrease of $112$ km~s$^{-1}$
  in the transition from \rm{SE1a} to \rm{SE1b} and a decrease of $36$ km~s$^{-1}$ in
  the transition from \rm{SE1b} to SE1c. 
  Further out (i.e. at larger distances from the central source) the observed
  radial velocities also decrease with distance. 

 Fig.~\ref{FigVel} 
 clearly illustrates the decrease of the radial velocities with 
  increasing distance to the source, and the change of $\sim300$ km 
  s$^{-1}$  in the radial velocity from the innermost regions to the 
  intermediate knots. The radial velocity also 
  decreases, although only by $\sim50$ km~s$^{-1}$, from the intermediate 
  knots \rm{NW2} and \rm{SE2} to the outermost \rm{NW3} and SW3, which are just grazed by 
the slit position  
  (see Fig.~\ref{FigVel}).  Several mechanisms can be proposed  
  to explain these variations in radial velocity 
  and they will be discussed later.

\subsection{Kinematics of the individual knots}
  
In the following we will analyse in detail the  internal kinematics of some 
of the individual knots. 
For this we will combine previously existing and new ground-based 
observations with the high resolution capabilities of STIS. 
  
\begin{figure*}
     \centering 
     \includegraphics[width=12cm]{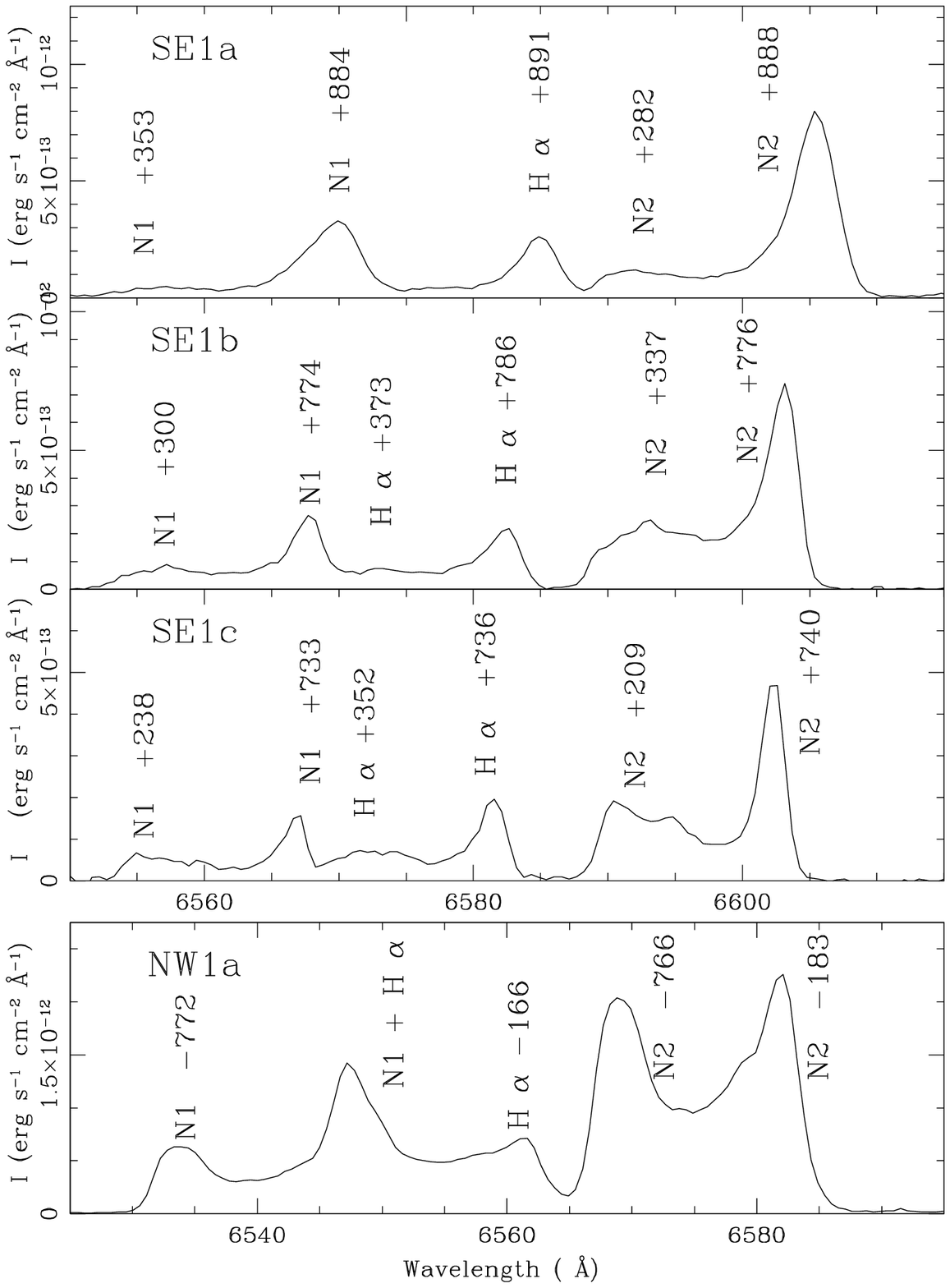}
     \caption{Intensity vs. wavelength profiles of [N~{\sc ii}] 6548,
  6584 \AA, and H$\alpha$ 6563 \AA~ for the innermost knots SE1a/b/c and
   \rm{NW1a}   derived by co-adding the signal over the corresponding 
  angular extent, obtained with  STIS at slit positions with
  offsets $-0\farcs1$, $0\farcs0$ and $+0\farcs1$ with respect to the central
  source. The vertical labels are the relative radial velocity of the peaks.
 The emission lines are also identified (N1 for [N~{\sc ii}] 6548 \AA, 
N2 for  [N~{\sc ii}] 6583  \AA). 
  }
      \label{FigSUM}
      \end{figure*}

Fig.~\ref{FigSUM} shows the emission line profiles derived from STIS data 
  of the [N~{\sc ii}] 6548, 6583 \AA, and H$\alpha$ 6563 \AA~ at the innermost 
  subcondensations \rm{SE1a} to \rm{SE1c} and  NW1a. The faint subcondensations \rm{NW1b}  and 
  \rm{NW1c}  were barely detected with STIS and consequently are 
  not shown here. The spectra shown are the result of 
  co-adding the signal over the entire spatial extent of each subknot, 
  at slit positions correspondings to the 
  offsets $-0\farcs1$, $0\farcs0$ and $+0\farcs1$ with respect to central 
  source (see Fig.~\ref{FigHST}). 
  
   The vertical labels are  the relative radial 
  velocities of the high- and low- velocity peaks for the compact 
  subcondensations SE1a, \rm{SE1b} and \rm{SE1c} in the receeding lobe, and for the knot \rm{NW1a}  
  in the approaching lobe (Fig.~\ref{FigSUM}). The wide range of 
  velocities covered by each of these compact subcondensations, from $37$ to 
  $1093$ km~s$^{-1}$ for SE1a, from $-23$ to $975$ km~s$^{-1}$ for SE1b, and from 
  $50$ to $865$ km~s$^{-1}$ for SE1c, are the largest observed in a PPN. 
  A similar situation 
  is observed in the blueshifted knot NW1a, where all the lines  
  have redward emission extending to $\simeq80$ km~s$^{-1}$.  
  
  The \rm{SE1} subcondensations show double-peaked emission profiles in all the 
  observed emission lines. 
  The strongest peaks are narrower than the weaker and wider low-velocity  
  maxima (see Figs.~\ref{FigSUM}). There is a hint 
  of a double-peaked profile in 
  SE1a, with the weaker component at approximately $280$ km~s$^{-1}$. The 
  double-peaked profiles are more pronounced in the emission lines formed in  
  \rm{SE1b} and SE1c. 

 The condensation \rm{NW1a}  differs from the previously discussed 
  subcondensations in that the low velocity component has slightly stronger  
  intensity than the high velocity component. The emission in \rm{NW1a}  can be 
  seen from about $-970$ km~s$^{-1}$ to $+80$ km~s$^{-1}$ 
  (see Figs.~\ref{FigSUM}). 
   Fig.~\ref{FigSTIS2} also shows that the
  emission extends beyond NW1a. Weak emission is detected 
  at a distance of $3\farcs2$ from the central core (close to NW1b), with 
  double-peaked  emission line profiles at  velocities of $-720$ and $-230$  
  km~s$^{-1}$ for the high- and low-velocity peaks, respectively. 

 The intermediate knots (\rm{NW2} and \rm{SE2}) show double-peaked profiles in all 
  lines. A detailed inspection of Fig.~\ref{FigVel}
  shows that the low velocity component of these double-peaked profiles 
  seems to linearly increase its velocity
  with increasing distance to the center while
  the high-velocity component (redward in \rm{SE2} and blueward in \rm{NW2}) remains
  almost constant. The velocity of the low component
  increases smoothly by $84$ km~s$^{-1}$ along $\sim 3''$ 
 in the redshifted lobe, 
  and $77$ km~s$^{-1}$ in $\sim 2''$ in the approaching lobe, while
  the high velocity components remain almost constant, 
  with fluctuations of just $20$ to $30$ km~s$^{-1}$.
  The mean outflow velocity of both knots is $455\pm15$ km~s$^{-1}$, 
  slightly higher than the value of $425$ km~s$^{-1}$ reported by R95. 
   The emission lines from these  knots show  extremely large widths of 
  $\sim475$ km~s$^{-1}$, slightly larger than the $425$ km~s$^{-1}$ derived
  from the emission further away (close to knots \rm{NW3} and \rm{SE3}). 
  It is noteworthy that the low-velocity peaks of \rm{NW2} and \rm{SE2} 
 knots are stronger  in intensity than the high-velocity ones.

  Fig.~\ref{FigSTIS1} shows the  velocity-position contour diagram of 
the lines H$\alpha$ and [N~{\sc ii}] 6584 \AA~  obtained from the STIS
 spectra of the \rm{SE2} knot (see Sect.  2.2.2). 
  We can see that the spectrum 
  displays a highly marked asymmetric shape with two intensity 
  maxima. This is very similar to the velocity structure usually observed in 
  the spectra of Herbig-Haro (HH, hereafter) objects (Raga 1995).

  The STIS spectrum of \rm{SE2} also shows that the two
  maxima are spatially displaced (one with respect to the other). 
  The high-velocity maxima occur $\sim0\farcs15$ 
  farther away from the central star ($1.3\times10^{16}$ cm at a distance of 5.8 
  kpc). The same was noticed to occur at knot \rm{SE1} by BH01, with the 
  lower velocity gas nearer the central star (by $\sim0\farcs05$).

Finally,  as we have already mentioned, for the first time we have also
 obtained spectra of the emitting gas of Hen~3$-$1475 beyond the position 
of the intermediate knots \rm{SE2} and \rm{NW2}. Our ground-based spectra show that
 at the outermost regions (i.e. at distances
  of $\pm 9''$ from the central source)  
  the emission lines show extremely broad, single-peaked  profiles with
  a mean outflow velocity of
  $\pm$ $405\pm15$ km~s$^{-1}$ relative to the systemic 
  radial velocity (see Table~\ref{tabradial}), 
  with a line width (i.e. FWHM) of $\sim425$ km~s$^{-1}$.

  \begin{table*}
  \centering
  \caption[ ]{Relative radial velocities of [N{\sc ii}] 6584 \AA}
  \begin{tabular}{lccccc}
  \hline
  Region & Observations & High & Low & Average & Line width  \\
         &              & (km~s$^{-1}$) &  (km~s$^{-1}$) &  (km~s$^{-1}$) &  (km~s$^{-1}$) \\
  \hline
  \rm{NW1a}  & HST + STIS  & $-$766 & $-$183 & $-$475 & 1050 \\
  \rm{SE1a} & HST + STIS  & $+$888 & $+$282 & $+$585 & 1056 \\
  \rm{SE1b} & HST + STIS  & $+$776 & $+$337 & $+$556 &  998 \\
  \rm{SE1c} & HST + STIS  & $+$740 & -- &  --  &  815 \\
  \rm{NW2} & INT + IDS & $-$552 & $-$367 &   $-$460 & 475  \\
  \rm{SE2} & INT + IDS & $+$540 & $+$355 &   $+$447 & 475 \\
  \rm{NW3} & INT + IDS &  --    &  --    &   $-$403 & 425  \\
  \rm{SE3} & INT + IDS &  --    &  --    &   $+$405 & 425   \\
 \hline
  \end{tabular}
  \label{tabradial}
  \end{table*}

  The mean relative radial  velocities of the blueshifted
  and redshifted peaks derived from our ground-based and STIS spectroscopy 
  are tabulated in Table~\ref{tabradial} for these three systems of 
  symmetric knots.   The estimated uncertainty for each of
  these individual velocities is $\pm15$ km~s$^{-1}$ 
for ground-based observations, and  $\pm25$ km~s$^{-1}$ for the STIS spectra. 
  The improvement of the  S/N ratio of these spectra with respect to previous 
  observations (e.g., R95)  allows a more accurate measure of the 
  differences between the low and high velocity emission peaks shown by
  these knots. These are marked as `low' and `high' in Table~\ref{tabradial}, 
 which also includes the average velocities and the line widths.

  \subsection{Proper motion measurements and distance}
  
  Based on a detailed comparison of the HST WFPC2 images obtained on 1996
  June 6 and 1999 September 19 through the F658N filter we have been able to
  derive for the first time the proper motions of all the system of knots and 
  subcondensations of Hen 3$-$1475 above described. BH01 reported 
  proper motion measurements only for the intermediate knots, since their 
  main interest was just the determination of the distance towards Hen
  3$-$1475. We have extended the analysis to 
  all the emitting knots in Hen 3$-$1475  because this can provide us relevant 
  information on the physical process(es) responsible for the formation of 
  these structures.

  \begin{table*}
    \caption{Proper motion measurements and tangential velocities in Hen~3$-$1475}
    \begin{center}
      \begin{tabular}{lcccc} \hline
       Knot  & $\mu$x & $\mu$y & Tangential velocity & P.A. \\ 
            & (milliarcsec yr$^{-1}$) & (milliarcsec yr$^{-1}$) &  (km~s$^{-1}$) & ($^{\circ}$) \\ 
\hline
         \rm{SE2} &  $-0.99\pm4.70$  & $+13.80\pm0.90$   & 377 & 140 \\
         \rm{SE1c} & $-4.35\pm4.82$  & \phantom{1}$-1.49\pm2.41$  & 125 & 245 \\
         \rm{SE1b} & $+0.69\pm4.60$  & $+13.80\pm0.90$  & 376 & 133 \\
         \rm{SE1a} & $-1.60\pm4.82$  & \phantom{1}$+1.00\pm1.14$  & $\phantom{1}$51 & 194 \\
         \rm{NW1a}  & $-1.34\pm4.56$  & \phantom{1}$-7.08\pm0.86$  & 196 & 305 \\
         \rm{NW1b}  & $-2.70\pm4.83$ & \phantom{1}$-3.86\pm2.60$  & 128 & 281 \\
         \rm{NW1c}  & $-0.04\pm4.96$ & \phantom{1}$-3.60\pm1.67$  & $\phantom{1}$98 & 315 \\
         \rm{NW2} &  $-2.89\pm4.73$ & $-12.60\pm0.86$  & 353 & 312 \\
         \rm{NW3} &  $-4.46\pm4.73$ & $-12.60\pm1.12$  & 365 & 297 \\ \hline
      \end{tabular}
      \label{tabprop}
    \end{center}
  \end{table*}
 
   In order to carry out these proper motion measurements,
  the first- and second-epoch F658N images were converted into a common
  reference system, with the Hen~3$-$1475 jet along the $y$ axis,
  oriented at a P.A. of $136^{\circ}$.  After the
  transformation, the average and r.m.s. of the difference in position for 
  eight reference stars in the field appearing in 
  two images was  $-0.318\pm0.080$ pixels 
  (equivalent to $-0\farcs0145\pm0\farcs0036$)  in the $x$ coordinate and 
  $0.050\pm0.060$ pixels (equivalent to $0\farcs0031\pm0\farcs0027$) in the $y$ coordinate.

   After this, we computed the two-dimensional cross-correlation function
  of the emission detected within small, previously 
  defined boxes containing the individual condensations under study.  Finally,
  the proper motions were determined through a parabolic fit to the peak of
  the cross-correlation function (see Reipurth et al. 1992 and L\'opez et al. 
  1996 for a detailed description of this method). The uncertainty in the 
  position of the correlation peak derived this way was estimated through 
  the scatter of the
  correlation peak positions obtained from boxes differing from the nominal
  one in 0 or $\pm2$ pixels ($0\farcs090$) in any of its four sides,
  making a total of  3$^4$ = 81 different boxes for each knot. The error 
adopted has been twice the r.m.s. deviation of the position found, for each 
  coordinate, added quadratically to the r.m.s. alignment error.
  The numerical results obtained for the system of knots in 
  Hen~3$-$1475 are shown in Table~\ref{tabprop}, which includes 
 the derived proper motions (in milliarcsec yr$^{-1}$), the tangential 
  velocities (in km~s$^{-1}$) for 
  an adopted distance of 5.8 kpc (see below), as well as the 
  position angle of the proper motion vectors (in degrees).
  
\begin{figure*}
     \centering 
     \includegraphics{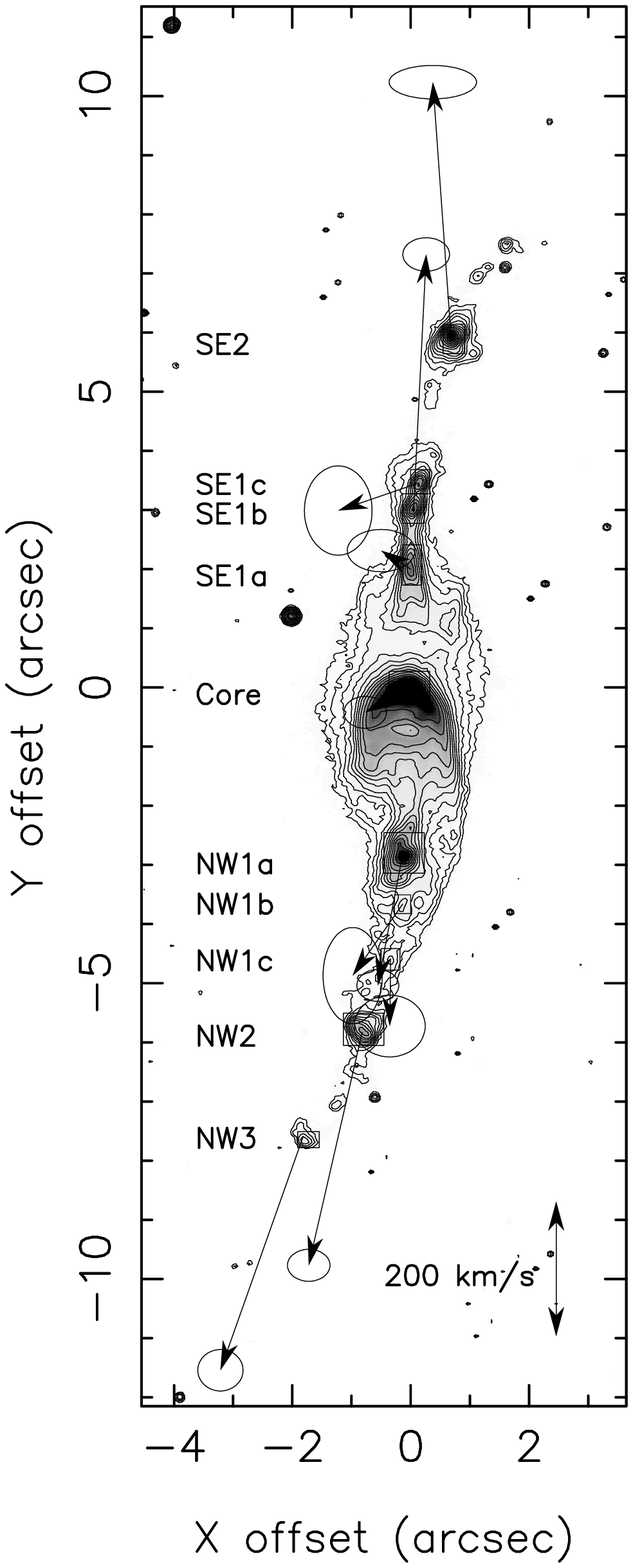}
     \caption{ HST image of Hen 3$-$1475 taken through the F658N filter derived from 
  the 1999 observations. The contour spacing is logarithmic. 
   The arrows indicate the proper motion
         velocity of each knot. The ellipses at the end of each arrow 
  indicate the uncertainty in the components of the velocity vector. 
  The (0,0) position was set at the location of the core, where the maximum 
  emission was measured. }
                \label{FigPROPER}
      \end{figure*}

 In Fig.~\ref{FigPROPER} we show the F658N HST 1999 image of Hen~3$-$1475,
 with each knot identified with a label. The arrows indicate the proper 
  motion velocity vector associated with each knot derived with this method, 
  and the ellipses at the end of each arrow indicate the uncertainty.
   
   In order to perform an accurate determination of the distance to Hen 
  3$-$1475, it is appropriate to base it on the proper motions of the 
  intermediate knots, which most likely correspond to 
  internal working surfaces (see Sect.  4), and whose  radial velocities  
  and proper motions should provide a good measure of the actual 
  space velocity of the knots (Raga et al. 1990).

   We find remarkably similar proper motions for the \rm{NW2} and \rm{SE2} knots,  
  with an average value of $13.2$ milliarcsec yr$^{-1}$, somewhat larger than 
  the results obtained by BH01. Combining the radial velocity of  $455$
  km~s$^{-1}$ associated to with these knots 
 (see Sect.  3.4) with an inclination 
  angle of the axis of $50^{\circ}\pm5^{\circ}$ with respect to the 
  plane of the sky (BH01), we find  the distance to Hen 3$-$1475 to be 
  5.8 kpc (for i = $45^{\circ}$ and $55^{\circ}$ the distances would be 
  6.9 and 4.9 kpc, respectively), significantly smaller than the value  
  of $8\pm2$ kpc derived by BH01. BH01 adopt a larger radial velocity and a 
lower proper motion for the intermediate knots resulting in a larger distance.

   In the \rm{NW} lobe, tangential velocities range from $100$ to $365$ km~s$^{-1}$ 
(for an adopted distance of 5.8 kpc). 
  The knots in this lobe move along the jet axis at an average position angle 
  of $302^{\circ}$. We find remarkably similar tangencial velocities 
   ($\sim350$ km~s$^{-1}$) and large proper motions for the knots \rm{NW2} and \rm{NW3} 
  while the subcondensation  \rm{NW1a}  shows a smaller tangencial velocity 
  ($\sim200$ km~s$^{-1}$) and proper motion in the direction of the 
  jet axis. Even smaller tangencial velocities and proper motions 
  are detected (just at  $\sim2\sigma$ level) 
  in the subcondensations \rm{NW1b}  and NW1c, although they still seem to 
  follow very closely the axis of the jet.
  
   In the SE lobe both the knot \rm{SE2} and the inner subcondensation \rm{SE1b} show 
  a remarkably similar tangencial velocity ($376$ km~s$^{-1}$) and 
  proper motion along the axis
  of the jet (with an average position angle of $136^{\circ}$). 
  However, the innermost knots \rm{SE1a} and \rm{SE1c} do not show any significant proper
  motion within the errors. The 
  measurements may be affected by the variation with 
 time of their shapes, as
  suggested by the completely different direction of motion shown by these
  subcondensations.
   Unfortunately, we could not determine the proper motion of knot \rm{SE3} 
  since this knot was affected by a bad CCD column in one of the images.

  Using the proper motions derived above for the knots 
  and their angular distances to the central source, we can estimate
  kinematic ages for the individual knots. The position of \rm{NW2} and \rm{SE2} 
  (quoted in Table~\ref{tabvel}) together with the derived
  proper motion of $12.6$ and $13.8$ milliarcsec yr$^{-1}$, respectively,   
  lead to a kinematic age of $\sim450$ years for the intermediate knots. 
  A similar procedure applied to the outermost knots \rm{NW3} and \rm{SE3}, 
  adopting for both the proper motion of $12.6$ milliarcsec yr$^{-1}$ 
measured for \rm{NW3}, gives a kinematic age of  $\sim600$ years. 
The kinematic ages of the intermediate and outermot knots (with an age difference of 
$\simeq150$ years between both system of knots) is compatible with the 
evolutionary stage of the central star and with the 
variable velocity model outlined below (see Sect. 4.3.3).  

  \begin{table}
  \centering
  \caption[]{An estimation of the kinematic ages of the intermediate and
  outer pairs of knots observed in Hen 3$-$1475}
  \begin{tabular}{lccc}
  \hline
  Knot & Angular  & Tangential & Dynamical \\
      & distance & velocity & ages \\ 
      &    (arcsec)      & (km~s$^{-1}$) & (years) \\ 
\hline     
    \rm{NW2} &  5.83  & 353$\phantom{^*}$  & 463 \\
    \rm{SE2} &  5.92  & 377$\phantom{^*}$  & 429  \\
    \rm{NW3} &  7.82  & 365$\phantom{^*}$   & 620  \\
    \rm{SE3} &  7.57  & 365$^*$  & 601 \\
  \hline
 $^*$ assumed value
  \end{tabular}
  \label{tabvel}
  \end{table}
 
  \section{Discussion}

  \subsection{Line excitation mechanism}
  
  The emission observed in the intermediate knots of Hen~3$-$1475 was found
  to be formed in the recombination region of shock waves (R95). The
  spectra of \rm{NW2} and \rm{SE2} are characterized by extremely large 
  [N~{\sc ii}]/H$\alpha$ ratios 
  ($\sim3.0$), and electron densities of $\sim3000$ cm$^{-3}$. 
  R95 qualitatively determined the shock velocity 
  needed to reproduce the observed [O {\sc iii}]/H$\beta$ and 
  [O {\sc i}]/H$\alpha$ ratios by comparison with plane-parallel 
  shock models from Hartigan et al. (1987), which adopt 
  solar abundances in their calculations. 
  The ratios observed in the middle knots were  compatible with 
  intermediate shock velocities of $\sim100$ km~s$^{-1}$. 
  
  In order to estimate the N/H relative abundance     
  R95 compared the [N~{\sc ii}]/H$\alpha$ ratios with the planar shock models 
  (for a shock velocity of $100$ km~s$^{-1}$) computed by Dopita et al. 
  (1984a), and derived  a  N/H abundance of 
  ($3.6\pm0.6$)$\times 10^{-4}$ assuming a linear dependence between 
  the [N~{\sc ii}]/H$\alpha$ ratio and the N/H relative abundance. 

    We can also determine the shock velocities able to
  reproduce the observed optical spectra by relaxing the  {\it a priori} 
  assumptions on the shock velocity and fixing instead 
  the gas abundances. None of the predictions of 
  plane-parallel shock models for different shock velocities 
  in the literature (e.g. Shull \& McKee 1979; Dopita et al. 1984a;  
  Hartigan et al. 1987) can reproduce the  
  [N {\sc ii}] 6548, 6583 \AA/H$\alpha$ and [N {\sc i}] 5200 \AA/H$\beta$ 
  emission line ratios observed in Hen 3$-$1475.
  Neither the bow-shock models available to now can predict the observed 
line ratios (Hartigan et al. 1987). None of these models are  applicable
 to PNe abundances, except the plane-parallel shock model computed 
by Meaburn et al. (1988) to reproduce the spectra of the outer shell 
of $\eta$ Carinae, which shows very high [N {\sc ii}] 6583 \AA/H$\alpha$ 
emission line ratios ($\sim 2.8$). Meaburn et al. (1988) assumed a N/H 
relative abundance of $10^{-3}$ and a shock velocity of 140~km~s$^{-1}$. 
Since none of the available grids of plane-parallel or bow-shock models 
are applicable to PNe, we have obtained a grid of (steady 1-D)
  plane-parallel shock models with the photoionization-shock code MAPPINGS
  Ic (Binette et al. 1985; Dopita et al. 1984b) adopting the mean Type I PNe
  abundances which are characterized by their high N/O abundance ratio
   (Kingsburgh \& Barlow 1994), and a pre-ionized gas with 
  $n_{\rm H} = 50$ cm$^{-3}$ (which fits the observed electron densities derived
   from the [S {\sc ii}] doublet lines). By comparing the observed emission
  line ratios for the intermediate knots (Table 4 of R95) with our 
  predictions of planar shock models, we can determine the velocity of the 
  shocks responsible for the observed spectra.
  
\begin{figure}
 \resizebox{\hsize}{!}{\includegraphics{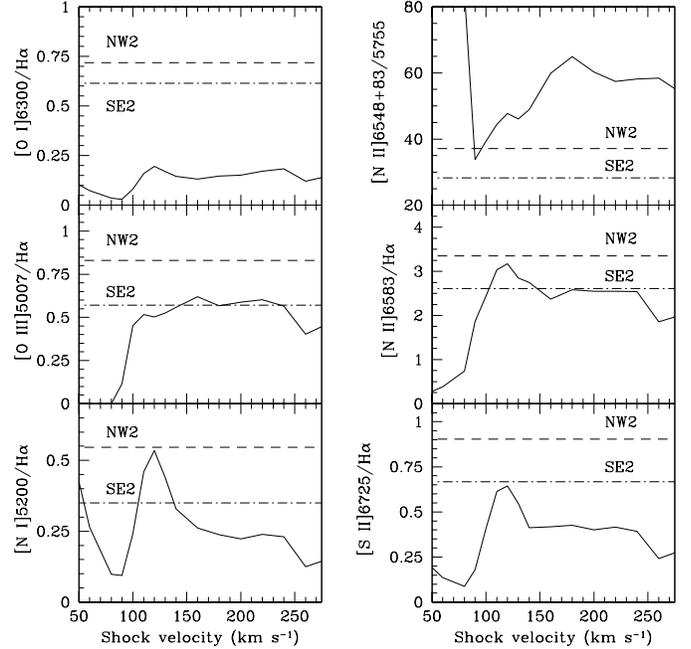}}
   \caption{Predicted emission line ratios for a grid of 1-D
  steady shock models as a function of shock velocity (solid line). Dashed
  lines are the observed emission line ratios for \rm{NW2} (dashed line) and \rm{SE2} 
  (dot-dashed line) (from R95).
  }
                \label{FigShock}
      \end{figure}

  The results are shown in Fig.~\ref{FigShock},
  where the observed and predicted emission line ratios are plotted
  for comparison. Fig.~\ref{FigShock} illustrates that the observed
  [N~{\sc ii}]~6584/ H$\alpha$ ratios are reasonably well reproduced for a 
  wide range of shock velocities
  (from 100 to 250 km~s$^{-1}$), while the values of the temperature-dependent
  ratio [N {\sc ii}] (6584 $+$ 6548)/5755 are only marginally reproduced
  for a shock velocity $\sim$ 100 km~s$^{-1}$ (other models 
  predict higher values which means lower T[N {\sc ii}]).
   The [N~{\sc i}]~5200/ H$\alpha$ ratios are reproduced for shock
  velocities from $100$ to $150$ km~s$^{-1}$. 
  These results confirm that the gas is nitrogen enriched. 
 
   In addition, the observed [O~{\sc iii}]~5007/ H$\alpha$
  ratios are reproduced only for shock velocities higher than $120$ km~s$^{-1}$ 
  (at least the observed value in the \rm{SE2} knot).
  
   However, the models predict too low [O~{\sc i}]~6300/ H$\alpha$ and
  slightly low [S~{\sc ii}]~(6717$+$6731)/H$\alpha$ ratios.
   The weakness of the [O~{\sc i}] and [S~{\sc ii}] emission lines predicted
  by the shock models, leading to a disparity between the observed and 
  predicted
  emission line intensities,  is not surprising since shock models appear to
  have problems in reproducing the observed [O~{\sc i}]/H$\alpha$ and
  [S~{\sc ii}]/H$\alpha$ line ratios 
  in SN remnants and HH objects too (see, e.g., Dopita et al. 1984b; 
  Raga et al. 1996).
  
   A clear problem is that we are comparing the emission
  line ratios from the knots \rm{NW2} and \rm{SE2}, which are curved structures, with
  plane-parallel steady models. In a curved shock we expect contribution 
  from the
  more oblique, lower shock velocity regions, and from higher shock velocity
  regions. The geometry of the condensation also influences on the ionization
   of the pre-shock gas, which strongly determines the post-shock emission. 
  On the other hand, the plane-parallel models presented here do not 
 include magnetic fields. We assume that the strength of the magnetic field 
is small at the location of the intermediate knots. If  
  magnetic fields ($\sim$ 0.1 to 30$\mu$G) are included 
the compression in the recombination region is reduced 
(Shull \& McKee 1979, Morse et al. 1992). The effect on the observed spectra is a 
reduction of the  [O~{\sc i}] and [N~{\sc i}] intensities (Shull \& McKee 1979, Morse et al. 1992, 
Morse et al. 1993). If the magnetic fields are strong we might be understimating the shock velocity 
(as pointed out by Trammell \& Goodrich 2002).

   In summary, the attempt to model the observed emission line ratios in
  Hen 3$-$1475 using plane parallel
  shocks can be considered as partially successful. The results are in most
  cases consistent with a spectrum arising from
   the recombination region of a shock wave with shock velocities ranging from
  $100$ to $150$ km~s$^{-1}$, moving through a nitrogen enriched gas.

  \subsection{Analysis of the emission line profiles}
  
  In the following we will try to describe the main properties of the
  line profiles observed in order to set constraints on the physical 
  mechanisms responsible for this emission. 
  
  Some of the properties of the emission knots in Hen~3$-$1475
  have been observed before in at least some well-known HH objects.  They 
  show also prominent double-peaked profiles and extraordinarily large line width.  
 For instance, the HH 32 objects show line widths exceeding $350$ km 
  s$^{-1}$ (Solf et al. 1986; Hartigan et al. 1986a); the HH objects in 
  Cepheus A show large line widths $\sim475$ km~s$^{-1}$ 
  (Hartigan et al. 1986b), while the most extreme radial velocities are
  observed in HH 80-A and HH 81-A, with a line width of $700$
  km~s$^{-1}$ (Heathcote et al. 1998).
  
  Usually, the low-velocity component of the double-peaked emission line 
  profiles observed in these HH objects is stronger than 
  the high-velocity component, as is also observed in  the
  intermediate knots of Hen 3$-$1475.    
  The reversed situation, which is found at the innermost subcondensations of
  SE1,  has also been found in a few HH objects, as HH 32D (Solf et al. 
  1986; Hartigan et al. 1986a).

   HH objects showing wide double-peaked emission line profiles 
  have been explained in terms of bow-shock models (e.g. Raga \& B\"ohm
  1986; Hartigan et al. 1987). The bow-shock contains a large range of 
  velocities projected along the line of sight that explain  how a small 
  volume can produce wide emission line profiles.  
  Under certain inclination angles (for oblique viewing angles), the 
  bow-shock models predict the existence of two maxima with the low-velocity 
  maximum being brighter than the high-velocity peak (``interstellar bullet'' 
  or bow-shock facing away from the central source)  or with the high-velocity 
  component stronger than the low-velocity emission  peak 
  (``stationary shocked cloudet'' or bow-shock facing towards the central 
  source). 
  
  The intermediate knots \rm{SE2} and \rm{NW2} in Hen 3$-$1475 share some of the 
  properties mentioned above for ``facing away bow-shocks": 
  extremely wide emission line profiles with the presence of two peaks, where
  the low-velocity component shows a higher intensity, and both knots
  show high proper motions. The inclination angles, 
  with respect to the plane of sky, required to reproduce the observed 
  double-peaked profiles (which have to be $\geq45^{\circ}$; Raga \& 
  B\"ohm 1986; Hartigan et al. 1987) are compatible with the adopted 
  inclination angle of $50^{\circ}\pm5^{\circ}$ for Hen 3$-$1475.   
  There are, however,  some dificulties with the simple bow-shock model. 
  The spatial velocity displacements of the low and high radial velocity gas 
  found in the 2-D spectra of Hen 3$-$1475 and the increase of the 
  the radial velocity of the low-velocity gas in \rm{NW2} and \rm{SE2} with distance to
  the source contradict the predictions of 2-D bow-shock models (Raga \&
  B\"ohm 1986). We should note that the details of the density structure in the shock has 
 a strong effect on the shapes of the emission-line profiles, even with 
 simple geometries.  Bobrowsky \& Zipoy (1989) show a variety of line 
 profiles arising from different density distributions along the line of sight. 

 In addition, the intermediate knots of Hen~3-1475 
 show high tangential  velocities comparable to the tangential velocities observed in 
 several HH objects (see, e.g., Eisl\"ofel et al. 1994), and in agreement
  with predictions of the variable ejection velocity  models (see below). 
 The complex structure and kinematics of the innermost knots remain 
 intriguing. These knots show extremely large line widths 
 ($\sim 1000$ km~s$^{-1}$, see Table~\ref{tabradial}) which imply 
 a complex kinematics on subarcsecond scales. The situation is apparently 
 more difficult to explain considering that subcondensations 
  \rm{NW1a}  and \rm{SE1b} 
 appear to move with high tangential velocities while subcondensations  \rm{SE1b} 
 and SE1c, and possibly \rm{NW1b}  and NW1c, appear to be stationary.  
  
  \subsection{Kinematics of the outflow}
  
  BH01 interpreted the emission line profiles of the knots as well as the 
  decrease of the radial velocity with distance to the source as signatures 
  of violent deceleration of the outflow both at small (a few $10^{16}$ cm) 
  and large (along the axis within a few $10^{17}$ cm) scales as the result of 
  turbulent interaction between the jet and the surrounding ambient gas.
  
   However, as we will see below, the strong deceleration claimed to occur 
  at small  scales within individual knots by BH01 (up to $\sim600$ 
  km~s$^{-1}$ in distances of the order of $\sim10^{16}$ cm)  cannot occur
  for realistic values of the relevant parameters in Hen 3$-$1475.  
   In addition, at large scales, there are  other mechanisms, apart from the 
  interaction of the jet with the surrounding ambient gas proposed by BH01, 
  which cannot be discarded as responsible for the observed decrease of
  the radial velocity, such as geometrical effects (due to precession) 
  and/or to variability of the ejection velocity (consequence of a variable,
  may be episodic, stellar wind), which are briefly discussed below.

  \subsubsection{A geometrical effect induced by precession}
    
   At first sight, the lack of alignment between the different knots seems 
  to favour the possibility that the direction of ejection could actually 
  be varying with time. In order to reproduce the observed 
  S-shaped morphology of Hen 3$-$1475 in terms of  precession, this motion 
  would require an aperture angle $<10^{\circ}$, 
  and a period $\sim1500$ years. However, such a precession, considered 
alone,  cannot reproduce the whole system of radial velocities observed 
  (see R95 for a more detailed discussion).
  
  \subsubsection{Interaction with the ambient medium}
  
  Two possibilities exist under this scenario. Either 
  the deceleration of the jet along its axis occur in a turbulent jet 
  beam which incorporates matter through a sideways entrainment process, or 
  this deceleration is the result of the interaction of a fragmented jet 
  with the external medium. 
  
  In order to explore the first of the two possibilities, 
  De Gouveia dal Pino (2001) simulated  a 3-D pulsed HH jet, illustrating  
  that  deceleration of the internal working surfaces (identified with the 
  knots)  occurs when moving away from the central source as momentum is 
  transferred by the gas expelled sideways from the traveling pulses.
  These simulations predict  
  a slow-rate decrease of the speed of the internal working surfaces, with 
   a final drop of the velocity at the position of the leading working 
  surface.   The smooth fall off of the velocity 
  of the internal working surfaces with distance (velocities decrease by 
  $\sim20$ km~s$^{-1}$ 
  along $5\times10^{17}$ cm for an initial ejection velocity of $\sim100$ km 
  s$^{-1}$) obtained by De Gouveia dal Pino (2001) do not fit the step-like 
  decrease of the  
  radial velocity with increasing distance and  the  abrupt velocity
  variation  at the position of the intermediate knots shown by Hen 3$-$1475. 
  
  This entraintment process could actually contribute to the decrease of the 
  velocity of the internal working surfaces as they travel away from the 
  central source, but cannot be responsible 
  for the bulk of the  slowing down observed along the jet of Hen 3$-$1475. 
  A numerical simulation with the actual 
  values (velocity, density, and time-scale 
  variability) applicable  for Hen 3$-$1475 would be needed to quantify the 
  exact contribution to the general slowing 
  down observed along the jet  that could be due to this effect.
  
   The second possibility is that the observed  deceleration could be 
  the result of drag forces of the ambient medium on individual jet knots, 
  as suggested by Cabrit \& Raga (2000) to explain the 
  jet deceleration in HH 34. This scenario is compatible with the 
  fragmentation into a series of clumps 
  of a jet with a periodic velocity variability and a slow precession 
  (Raga \& Biro 1993), which is likely to be the case of Hen 3$-$1475. 
  
  We  explore this situation, in which the jet fragments produce  
 ``interstellar bullets" 
  which are slowed down by the interaction with the surrounding material 
  (i.e. the AGB remnant) using the analytical 
  approach of Cabrit \& Raga (2000).  
  
  For this calculation we will choose a value of the isothermal sound speed 
  of $10$ km~s$^{-1}$. 
 In addition, we need to estimate the mass of the clumps and
  the environmental density, which are not well known.   
  If we consider a pre-shock density  of n$_{\rm H} = 50$~cm$^{-3}$ (see Sect.  
  4.1) and a size of a few $10^{16}$ cm for the 
  intermediate knots (adopting a distance of 5.8 kpc and a filling factor 
  $\sim$1), we can infer a mass for these knots of $\sim10^{-5}$ $M_{\sun}$.

  \begin{figure}
  \resizebox{\hsize}{!}{\includegraphics{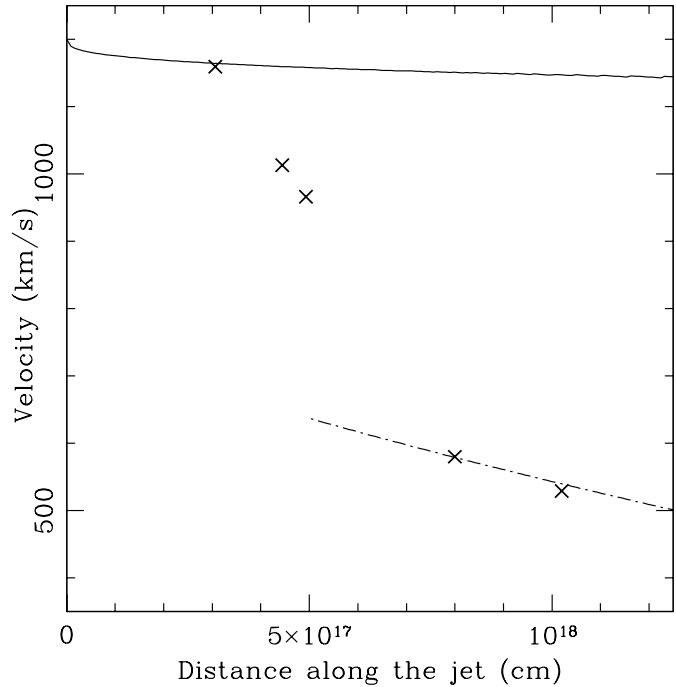}}
     \caption{Predicted deceleration curves (position-velocity) for 
  constant mass knots decelerated by the ambient medium (see Sect.  4.3.2). 
The values of the observed velocity and distance to the source of the 
innermost and intermediate knots are shown as crosses.}
                \label{FigDrag}
      \end{figure}
 
   With respect to the enviromental density, we have to consider 
  the presence of the  molecular  torus around the central source detected in 
  CO (Knapp et al. 1995; Bujarrabal et al. 2001) and OH (Bobrowsky et al. 1995;
  Zijlstra et al. 2001) which indicates the existence of an equatorial 
  density enhancement in the AGB ejecta. 
  
  Adopting the mass loss rate and the polar dependence of the AGB 
  wind from Mellema (1995), 
  a value $\leq300$ cm$^{-3}$ at a distance of $2.5\times10^{17}$ cm from the 
  central star is inferred 
  for the density of the environmental material that would interact with the 
  highly collimated gas ejected along the poles (for a ratio between the 
  density at the equator and at the poles of 5.0 and assuming that the environmental density
  decreases with distance as $r^{-2}$).   
The  deceleration curve (position versus velocity) 
  derived is shown as a   solid line in Fig.~\ref{FigDrag} for an 
 initial velocity of $1200$ km~s$^{-1}$.  
   We can see that the predicted decrease of the radial 
  velocity with distance to the source due to the drag force cannot explain 
  by itself the strong deceleration observed in the innermost regions of the
   outflow, nor  the difference in 
 velocity observed between the internal 
  knots  and the intermediate ones. 
  However,  it can perfectly explain the observed deceleration from \rm{SE2} to \rm{SE3} 
  (dotted line of Fig.~\ref{FigDrag}) 
  for an initial velocity of $\sim775$ km~s$^{-1}$ and a density of 
  $\sim50$ cm$^{-3}$ at the location of the intermediate knots.  
  
   Our conclusion is that only the small deceleration occuring at large scales
  (slowing down of the velocity of the knots of $\sim50$ km~s$^{-1}$
  along few $10^{17}$ cm) from the intermediate knots to the outermost knots
  of Hen 3$-$1475 can be explained as the result of drag forces of the 
  ambient medium and/or entrainment as the knots move outwards,  while the 
  strong deceleration claimed to occur  at smaller scales by BH01 
(with deceleration of $\sim600$ km~s$^{-1}$ 
  along few $10^{16}$ cm) are unlikely to 
  be caused by this environmental drag. 

  \subsubsection{A changing stellar wind with variable ejection velocity}
  
  The step-like velocity variations observed in the innermost subcondensations
  very close to the central source in Hen 3$-$1475 invite to suggest the  
  possibility of interpreting these knots                   
  as bow-shocks formed in the outflow as a consequence 
  of  a changing stellar wind with a variable velocity ejection (perhaps episodic mass loss events). 
  
  Different physical scenarios might form bow-shocks. In the
 ``shocked cloudet''
  model the HH knots are slowly moving condensations, embedded in a
  supersonic wind ejected from the exciting source (Schwartz 1978). 
  Bow-shocks  directed away from the central star can be formed by
  the interaction of clumps of gas  with the ambient medium
  during  discrete events of ejection (scenario known as the ``interstellar bullet'' model 
  which was introduced by Norman \& Silk (1979)).  
  More recent models are based on the idea that HH emission is associated with 
  shocks produced in jet-like outflows. Bow shock-like structures (identified with the knots) 
 can be formed in a   jet-like outflow as a result of a variability of the ejection velocity  
   (see Raga 1991 for a detailed description 
  of the origin of the bow-shock like structures within HH jets; Hartigan \&
   Raymond 1993; Stone \& Norman 1993a,b; De Gouveia Dal Pino \& Benz 1993, 
  1994). It has been shown that 
  for a quasi-periodic velocity variability with a highly supersonic velocity
   amplitude, shocks develop  in the beam which are  arranged in two-shock 
  structures called `internal working surfaces` which are identified with the 
  knots (see, e.g., Raga et al. 1990; Hartigan and Raymond 1993).

  Raga \& Biro (1993) showed --- analytically and numerically --- that a jet with a 
  quasi-periodic variability and a slow precession would fragment into clumps
   which behave like the ``interstellar bullets" proposed by Norman \& Silk 
  (1979). Such a variability might be responsible for the formation of the 
  knots of the jet in Hen 3$-$1475, since its point-symmetric morphology 
  suggests that the direction of ejection is actually varying as a function 
  of time. The predictions of this model are compatible with the high 
  proper motions measured for the knots, the presence of the double-peaked 
  wide emission line profiles,
  the relative intensities of the high and low-velocity components (at least 
  in the intermediate knots) and the shock velocities derived in Hen 3$-$1475.
   
   Detailed numerical models able to predict the kinematics 
  and emission line ratios along jets of this kind are needed
  to check whether other properties shown by Hen 3$-$1475 could also be
  explained in this scenario.  The extreme properties shown by the 
  innermost knots and the lack of proper motions in their subcondensations, 
   for instance, are difficult to reproduce with the simple 
  model outlined above.    
  
   To investigate whether a variation in the ejection 
velocity could induce the 
  formation of the knots observed in Hen 3$-$1475 we can carry out a
  comparison of some observed properties with the predictions of the 
  analytical 1-D model by Raga et al. (1990). 
  The 1-D analytical approximation for preassure-matched jets predicts that  
  the emission observed in the internal working surfaces  would mainly 
  come from shocks of velocity V$_s\simeq1/2(u_1 - u_2)$, and 
  the  internal working surface should be moving at
  $V_{ws}\simeq1/2(u_1 + u_2$), where $u_1$ and $u_2$ are the velocity
  of the flow before and after the jump associated with the knots (Raga et al.
   1990). From the data shown in Table~\ref{tabradial} for \rm{SE2} and \rm{NW2},  we deduce 
  shock velocities of $200$ km~s$^{-1}$, and tangential velocities for the 
  internal working surfaces  
  of $475$  km~s$^{-1}$ (for an inclination angle of 50$^{\circ}$). 
  The predicted V$_s$ is in qualitatively agreement with the observed spectra,  
  and   the observed 
 tangential velocities of \rm{NW2} and \rm{SE2} 
  ($\sim400$ km~s$^{-1}$) are also comparable to the predicted value of $475$  
  km~s$^{-1}$. It is interesting to notice the good agreement between the 
  predicted and observed values of the shock velocity and the velocity of the 
  intermediate knots from a simple 1-D model for a constant density jet.  
  
   For a highly supersonic flow (i.e. in the absence of preassure forces), the fluid parcels 
 move ballistically, and  for the asymptotic regime of large distances from the 
  central source, all the internal working surfaces should move with the same velocity 
  $V_{ws}$ (see Sect. 3.4 of Raga \& Kofman 1992). 
The variability period can then be calculated 
  by combining this velocity $V_{ws}$ with the spatial separation between two 
  successive knots (Raga \& Kofman 1992).  The intermediate and outermost
   knots of the blueshifted side of the jet move with similar 
  velocities as required above (see Table~\ref{tabvel}). From the spatial distance 
  between the intermediate and outermost knots (d$_{ws}$ taken from Table~\ref{tabvel}) 
  we estimate a timescale $\tau\simeq d_{ws}/V_{ws}\simeq 100$~years 
   for the variability of the velocity of the jet. 
  
  It is interesting to note that 
  variability over periods of few hundred years have been  found to occur in 
  some PPNe and PNe, through the detection of concentric circumstellar
  arcs and rings, viewed in scattered light (see Hrivnak et al. 2001 
  and references therein). 
  The arcs are interpreted as the result of
  quasi-periodic mass-loss events of very short duration experienced by the
  central star.  
  
   The physical origin of these short periods of enhanced mass loss 
  is still unknown. The associated timescales are much longer than the 
  observed pulsational periods of AGB stars (300 to 1000 days, typically), 
  but  much shorter than the timescale of thermal pulses ($10^4$--$10^5$
  years). 
  
  Several mechanisms have been proposed:
  \begin{itemize}
  \item 
  instabilities in the gas-dust coupling in a radiation pressure-driven
  outflow that can create density waves that appear as discrete shells
  (Deguchi 1977),
  \item 
  the possible influence of the presence of a binary companion, creating 
  cyclical perturbations in the mass outflow (Harpaz et al. 1997; 
  Mastrodemos \& Morris 1997),
  \item 
  a magnetic cycle leading to coronal ejection-like events 
  as a consequence of the 
  dynamo action induced by the differential rotation 
  between the rapidly rotating stellar core and the more slowly rotating
  envelope of the star (Soker 2000; Blackman et al. 2001a).
  \end{itemize}

   On the other hand, the apparent slowing down of the jet of Hen 3$-$1475 
(from $\sim1200$  km~s$^{-1}$ at the innermost knots to $\sim600$   
km~s$^{-1}$ at the intermediate knots)  would indicate that, superimposed on the 
quasi-periodic velocity variability giving rise to the internal 
  working surfaces (e.g. the knots), there is an underlying trend of 
  increasing ejection velocity over time (as was first proposed by R95).

  \subsection{Nature and evolutionary stage of Hen~3$-$1475}
  
  At a distance of 5.8 kpc the stellar luminosity of the central source of 
  Hen 3$-$1475 is $\sim12600 L_{\sun}$, typical of a relatively 
  high luminosity post-AGB star, but half the value 
  derived by BH01,  confirming the results of R95. 
  
   The mass loss rate can be estimated in the following way.
  If we idealize the flow as a cylinder of radius $R$, with gas of uniform 
  density $\rho$ that moves at a velocity $V$, then 
   $\dot{M}\simeq\pi\,R^2\,\rho\,V$.
  For the observed parameters at the position of the intermediate knots, with
   $R\simeq3\times10^{16}$~cm, a pre-shock density $\sim50$ cm$^{-3}$ and a 
  velocity of $600$ km~s$^{-1}$   (adopting an inclination of $50^{\circ}$), 
  we derive a mass loss rate $\dot{M} \sim2\times10^{-7} M_{\sun}$ yr$^{-1}$; 
also compatible with the star being in the post-AGB phase.  
  
  Therefore,  the wind  kinetic luminosity 
  and kinetic momentum  are $L\simeq\dot{M}\,V^{2}\simeq6\times10^{34}$ 
   erg s$^{-1}$ and $P\simeq\dot{M}\,V\,\tau \simeq10^{37}$ g~cm~s$^{-1}$
  (adopting a kinetic age $\tau\simeq600$ years), respectively. From the 
  above data, we  
  can also estimate the total kinetic energy of the high-velocity  outflow, 
  which is  $\simeq 10^{45}$ erg.

  Different models for the formation of highly-collimated jets in post-AGB 
  stars have been proposed.  Magnetohydrodynamical models can form 
 collimated structures  from a single rotating star 
 (Garc\'\i a-Segura et al. 1999), while 
  other theories require the presence of an accretion disk 
(Soker \& Livio 1994; Blackman et al. 2001b). Images of  
 the young PN He~3$-$1357 show collimation of outflows by an outer, previously 
 ejected gas shell (Bobrowsky et al. 1998).
  
  3-D MHD simulations by Garc\' \i a-Segura \& L\'opez
  (2000) showed that jets form as a result of the action of the hoop stress 
  caused by the toroidal magnetic 
  field; these jets can be detected if the mass loss rate of the fast wind 
  is $\geq 10^{-7} M_{\sun}$ yr$^{-1}$, which is the case of  
  Hen 3$-$1475.
  
  In the numerical simulations by Garc\' \i a-Segura et al. (1999) and 
  Garc\' \i a-Segura \& L\'opez (2000), jets 
 form at distances 
  of $\sim 10^{17}$ cm from the central star. 
  We should note that in  the simulations by Garc\' \i a-Segura \& L\'opez 
 the fast wind velocities are $\sim100$ km~s$^{-1}$,  
  which is significantly slower than the wind velocity observed in 
  Hen 3$-$1475 (R95; S\'anchez Contreras \& Sahai 2001). 
  The increase of the fast wind velocity in the MHD simulations of 
 Garc\'\i a-Segura et al. 
  does not prevent the formation of the highly collimated structures.   
  The only difference is that for higher wind velocities the collimation 
 of  the jet occurs at   even larger distances from the central source 
 (i.e. $>10^{17}$ cm).  
   This is more than the distance at which collimation is present in Hen 3$-$1475 
(at $\sim10^{16}$ cm; S\'anchez Contreras \& Sahai 2001).  
Otherwise, the mass loss rate and/or the magnetic fields  
  should be increased for the collimation to occur at shorter distances from 
  the central source. 
  
  In the MHD models, point symmetric morphology  
 may  be caused by  
  precession  or may result from a steady misalignment of the magnetized wind 
  axis with respect to the axis of the bipolar wind outflow 
  (Garc\' \i a-Segura 1997, Garc\' \i a-Segura \& L\'opez 2000). Both cases 
  require the existence of a binary companion.

  In order to explain the presence of episodic ejection (or quasi-periodic 
  variability) MHD instabilities in a magnetized outflow are invoked. The 
  3-D MHD simulations by 
  Garc\' \i a-Segura (1997) show that the collimated outflow is likely to 
  be subject to kink 
  instabilities, which can lead to the formation of blobs resembling the 
  knots of a jet.  Otherwise, the formation of an episodic jet can be produced 
by periodic variations of the magnetic field, as the the cyclic 
polarity inversion of the magnetic field  introduced  by 
Garc\' \i a-Segura et al. (2001) to reproduce
the concentric rings observed in many PPNe and PNe.  

  Another mechanism that might be responsible for the formation of the jet of 
  Hen 3$-$1475  is the presence of an accretion
  disk (Soker and Livio 1994; Blackman et al. 2001b). 
 The formation of an accretion disk  in PNe 
  requires the existence of a binary companion 
 (Soker 1998; Soker \& Livio 1994; Reyes-Ruiz \& L\'opez 1999) .

   The expected properties of a jet driven by 
  an accretion disk formed at the end of the common-envelope phase deduced 
by  Soker \& Livio (1994)  (mass loss rate into the jet  
  $\sim 10^{-7} M_{\sun}$ yr$^{-1}$ to  $10^{-6} M_{\sun}$ 
  yr$^{-1}$, and jet velocities $\sim 500$ km~s$^{-1}$)
 qualitatively agree with  the properties of the jet observed in Hen 3$-$1475. 
  
  Recently, Blackman et al. (2001b) present a MHD model to describe the 
  interplay between disk winds and stellar winds in the context of PPNe, where 
  both the star and the accretion disk may blow outflows. 
  For ages of a few hundred years, disk winds --- which dominates 
 over the stellar wind ---
 can supply $\sim2\times10^{34}$ erg s$^{-1}$, 
  which is compatible with the estimated kinetic luminosity
  ($\sim6\times10^{34}$ erg s$^{-1}$) and the estimated age ($\sim600$ years) 
 of Hen 3$-$1475.

  If the outflow is due to the magneto-hydrodynamical collimation of the disk 
  (or disk + star) wind, precession could be radiation-induced by 
  the irradiation of the disk by
   the central source (Livio \& Pringle 1997). As the disk wobbles, 
there are episodes of an 
 enhanced accretion rate with timescales $\leq1000$ years 
 (Livio and Pringle 1997),  that would produce an episodic jet.

Recently, the magneto-hydrodynamical collimation has been 
reinforced by the detection of the magnetic field in the torus of the 
PN K3$-$35 (Miranda et al. 2001), which shows strong 
similarities with Hen~3$-$1475. 
It has a  bipolar precessing jet and a dense equatorial toroid, and its 
jet is a strong source of [N~II] emission (Miranda et al. 2000).

  \section{Conclusions}
  
   The luminosity ($\sim12600 L_{\sun}$) derived assuming that the star is located at a distance of
  5.8 kpc, the nitrogen overabundance
  derived from the spectroscopic data,  the timescales estimated for  the
  development of the observed morphology and the rest of the 
  parameters analysed in this paper are all consistent with the 
  classification of Hen 3$-$1475 as a relatively massive star
  (M$\sim3-5 ~M_{\sun}$) in the post-AGB phase, as previously 
  suggested by R95.

 One of the most remarkable results found in the analysis of this flow 
is the variation of the radial velocity along the axis of the jet (see Sect. 3.3). 
 We have shown above that 
 the overall trend of the radial velocity with distance to the source 
 can be explained as a result of a variable velocity ejection 
 (maybe episodic mass loss events) 
 together with a slowing down with distance along  
 large scales due to the  entrainment process and/or as a result of the drag 
 forces of the ambient medium. 

 A scenario with a quasi-periodic variability (with timescales of $\sim100$ years) 
 and a slow precession (with a period of $\sim1500$ years) would explain the 
 formation of the knots, the observed morphology and the observed spectra 
 of Hen~3$-$1475.  Superimposed on the quasi periodic  velocity 
 variability there is an underlying trend of increasing velocity over time.  
 The ejection velocity variability predicts a shock velocity 
 which is significantly  lower than the velocity obtained for the 
 intermediate  knots, as required to explain the spectra observed in these 
knots. 

 The periodic variability is  most probably the consequence of 
 quasi-periodic mass-loss events of very short duration experienced by the
  central star, of still unknown origin.  In order to explain the observations, 
we can invoke instabilities in the mass outflow,
 the possible influence of a binary companion or coronal mass ejection-like
  induced by the magnetic field generated as a consequence of the dynamo
  action induced by the differential rotation 
  between the rapidly rotating stellar core and the more slowly rotating
  envelope of the star.

 The point-symmetric geometry displayed by Hen 3$-$1475 is most probably 
 caused by the precession  of the jet.  It has been shown by different 
 authors that precession requires the existence 
 of a binary system, 
 in which the wind, either from the central post-AGB star or from an 
 accretion disk is magneto-hydrodynamically  collimated. Otherwise, the 
 precession could also be radiation-induced by the irradiation of the disk. 
    
 The emission lines of several knots show double-peaked 
profiles  in both the intermediate and innermost knots,  with a line width
 from $\sim475$ km~s$^{-1}$  up to $1000$ km~s$^{-1}$ (see Sect. 3.4). 
 The huge dispersion of velocities observed in these knots were 
 interpreted as violent deceleration occurring in the knots by BH01. 
 However, we have shown that these  decelerations are unlikely to be 
caused by enviromental drag.  
 In the scenario we propose here,  these 
 knots are bow shock-like structures 
 which contain a large range of velocities projected along the line of sight. 
 In this way, we can explain how shuch an extremely wide double-peaked emission 
line profiles can be produced in a small volume.

  \begin{acknowledgements}
  AR is very grateful to E. de Gouveia Dal Pino, G. Garc\' \i a-Segura, A.
   Raga and L. Binette for fruitful discussions and comments. 
 We are grateful to 
  Luc Binette for providing us the photoionization-shock code MAPPINGS Ic. 
   AR and RE are partially supported by  DGICYT grant PB98-0670 and by MCyT grant 
AYA2002-00205. MB was  supported by STScI grant number GO-06364.03-A.
  This work was partially funded through grant PB97-1435-C02-02  from the 
  Spanish Direcci\'on General de Ense\~nanza Superior (DGES).
  \end{acknowledgements}
  
  {}
  \end{document}